\documentclass[12pt]{article} 
\usepackage{amsmath} 
\usepackage{amssymb} 
\usepackage{graphicx} 
\usepackage[letterpaper]{geometry}
\newcommand{\unity}{{\bf 1}\hspace{-1mm}{\bf I}}

\newcommand{\bpsi}{\mbox{\boldmath ${\psi}$}}
\newcommand{\bsigma}{\mbox{\boldmath ${\sigma}$}}

\newcommand{\odis}{\big<\hspace{-1mm}\big<}
\newcommand{\cdis}{\big>\hspace{-1mm}\big>}

\newcommand{\oter}{\big<}
\newcommand{\cter}{\big>}

\newcommand{\ceff}{\big>_\star}

\begin{document}

\title{ {\bf A spherical Hopfield model}}
\author{ D. Boll\'e$^a$, Th.M. Nieuwenhuizen$^b$, I. P\'erez Castillo$^a$, 
T. Verbeiren$^a$}
\maketitle
\begin{center}
$^a$Instituut voor Theoretische Fysica, Katholieke Universiteit Leuven,\\
Celestijnenlaan 200 D, B-3001 Leuven, Belgium\\
$^b$Institute for Theoretical Physics, Valckenierstraat 65, 1018 XE     
Amsterdam,\\ The Netherlands\\
\end{center}

\begin{abstract}
We introduce a spherical Hopfield-type neural network involving neurons 
and patterns that are continuous variables. We study both the 
thermodynamics and dynamics of this model.  In order to have a retrieval
phase a quartic term is added to the Hamiltonian.  The thermodynamics of
the model is exactly solvable and the results are  replica symmetric. A 
Langevin dynamics leads to a closed set of equations for the order 
parameters and effective correlation and response function typical for 
neural networks. The stationary limit corresponds to the thermodynamic 
results. Numerical calculations illustrate our findings.
\end{abstract}

\section{Introduction}
In general, the introduction of continuous versions of discrete spin 
models in statistical mechanics has always been welcomed. 
Since the work of Kac and Berlin \cite{KB52} the spherical model has 
lead to a better understanding of a lot of basic phenomena due to the 
fact that they simplify the mathematical calculations considerably. In 
spin glass research there is no doubt that an important role has been 
played by the spherical p-spin spin glass model \cite{CS92} \cite{Cr93}. 
Since then it has been quite remarkable to see how a mean-field model 
with spherical spins resembles  the results of realistic structural 
glasses (see, e.g., the many references in \cite{CrFe02}). 

In neural network theory  the Hopfield model is by now a well known 
{\it classic}  found in many textbooks (e.g., \cite{HKP91}, \cite{MRS95}) 
to study storage and retrieval in associative memory problems. Strangely 
enough, a spherical version of it has not yet been studied in the 
literature. One of the reasons might be  that by relaxing the neurons 
from Ising-type to spherical-type variables the retrieval phase is 
absent, i.e. one of the most important properties of this model is lost.

In this work we show that this problem has an easy and elegant solution. 
By introducing a quartic potential term in the Hopfield Hamiltonian such 
that the neurons stay spherical the retrieval phase is recovered. In this
way we obtain a simplified version of the Hopfield model with an exactly
solvable thermodynamics  leading to marginally replica symmetric results.  
A phase diagram is obtained with an explicit solution for the spin-glass
retrieval `transition' line, showing no re-entrance. Moreover, the region 
of global stability for the retrieval solutions is larger than the 
corresponding region in the standard Hopfield model.

Next, we study the relaxational Langevin dynamics of this model leading to
a closed set of equations for the order parameters and effective correlation
and response functions, typical for neural networks. These equations are 
similar to those of other models with continuous spins (see, e.g, 
\cite{Cu02} and references therein). We discuss the evolution of the overlap
order parameter in the retrieval phase. The stationary limit of the dynamics
is found to correspond to the thermodynamic results. The numerical solution
of the dynamics illustrates this behaviour.

The rest of the paper is organized as follows. In  section 2 we introduce 
the model. Section 3 contains the thermodynamical properties with the 
temperature-capacity phase diagram. In section 4 we study the dynamics 
showing the evolution of the retrieval overlap.  Finally, section 5 
summarizes our conclusions.

\section{A spherical Hopfield model}
The Hopfield model \cite{H82} is defined through the following mean-field
Ising-type Hamiltonian
\begin{equation}
{\mathcal H}(\{\bsigma\})=
             -\frac{1}{2}\sum_{i\neq j=1}^NJ_{ij}\sigma_i\sigma_j \, ,
\label{hopham}
\end{equation}
where the couplings $J_{ij}$ are related with the information one wants to
store in the network through the Hebbian rule
\begin{equation}
   J_{ij}=\frac{1}{N}\sum_{\mu=1}^p\xi^\mu_i\xi^\mu_j \, ,
\end{equation}
with $p=\alpha N$, where $\alpha$ is the loading capacity of the network. 
In this work, the $p$ patterns $\xi^\mu_i$ are choosen to be a collection 
of continuous indepnedent identical random variables (i.i.d.r.v.) with 
respect to $i$ and $\mu$ and drawn from a gaussian distribution
\begin{equation}
P(\xi^\mu_i)=
   \frac{1}{\sqrt{2\pi}}\exp\Big[-\frac{(\xi^\mu_i)^2}{2}\Big]
\end{equation}
and the neurons (spins) are also taken to be continuous and to satisfy the
spherical constraint
\begin{equation}
    \sum_{i=1}^N\sigma^2_i=N\, .
\end{equation}
As we will show explicitly below this setup of the spherical Hopfield model
does not allow for a retrieval phase. 
Therefore, we add the following term to the Hamiltonian (\ref{hopham})
\begin{equation}
-\frac{u_0}{4}\sum_{i,j,k,l}
          J_{ijkl}\sigma_i\sigma_j\sigma_k\sigma_l,
  \quad J_{ijkl}=
 \frac{1}{N^3}\sum_{\mu=1}^p\xi^\mu_i\xi^\mu_j\xi^\mu_k\xi^\mu_l \,, 
\end{equation}
which is a quartic term in the order parameter $m$ characterizing the 
retrieval phase 
\begin{equation}
     m^{\mu}=\frac{1}{N}\sum_{i=1}^N\xi^{\mu}_i\sigma_i \,.
\label{overlap}
\end{equation}

The introduction of this quartic term turns out to contribute 
macroscopically to the condensed part of the free energy but it only leads
to sub-extensive contributions to the noise produced by  the non-condensed
patterns and to a sub-extensive contribution coming from the diagonal 
terms, as we now discuss in the following section.

\section{Thermodynamic and retrieval properties.}
We apply the standard replica technique \cite{MPV87} in order to study the 
thermodynamical properties of the model defined above. Starting from the 
Hamiltonian  
\begin{equation}
{\mathcal H}(\bsigma)=
  -\frac{1}{2}\sum_{i\neq j}J_{ij}\sigma_i\sigma_j
   -\frac{u_0}{4}\sum_{i,j,k,l}J_{ijkl}\sigma_i\sigma_j\sigma_k\sigma_l
\label{Hamiltonian}
\end{equation}
and assuming, for simplicity, one condensed pattern, the replicated free 
energy per site becomes
\begin{equation}
\begin{split}
\beta f &= \lim _{N \to \infty} \lim_{n \to 0} \frac{-1}{Nn}
\int_{-\infty}^{\infty}\prod_{\alpha=1}^n dm^\alpha
\int_{-\infty}^\infty \prod_{\alpha\neq\beta}dq_{\alpha\beta} 
\int^{i\infty}_{-i\infty} \prod_{\alpha=1}^n\frac{du_\alpha}{4\pi i}
\int^{i\infty}_{-i\infty} \prod_{\alpha\neq\beta}
    \frac{d\widehat{q}_{\alpha\beta}}{4\pi i/N} \,\, \mbox{e}^{-Ng}
\end{split}
\end{equation}
with
\begin{equation}
\begin{split}
g&=-\frac{\beta}{2}\sum_{\alpha=1}^nm_\alpha^2
   -\frac{u_0\beta}{4}\sum_{\alpha=1}^nm_\alpha^4
    +\frac{n\alpha\beta }{2}
    +\frac{1}{2}\sum_{\alpha,\beta=1}^n m_\alpha(q^{-1})_{\alpha\beta}
         m_{\beta}\\
&-\alpha n \ln 2\pi
    +\frac{\alpha }{2}\ln\det({\unity-\beta{\bf q}})
     -\frac{n}{2}\ln2\pi+\frac{1}{2}\ln\det\widehat{{\bf Q}}
     -\frac{1}{2}\sum_{\alpha,\beta=1}^n\widehat{Q}_{\alpha\beta}
           q_{\alpha\beta} \,.
\end{split}
\end{equation}
In this expression 
\begin{equation}
  m_{\alpha}^1=\frac{1}{N}\sum_{i=1}^N\xi^1_i\sigma_i^{\alpha} \,, \quad
   \quad
   q_{\alpha \beta}= \frac1N \sum_{i=1}^N \sigma_i^\alpha \sigma_i^\beta,
  \quad \alpha \neq \beta
\end{equation}
are the retrieval overlap, respectively the neuron (spin) overlap order 
parameters, $\widehat{q}_{\alpha\beta}$ are conjugate variables and
\begin{equation}
\widehat{Q}_{\alpha\beta}=
          u_\alpha\delta_{\alpha\beta}
           -\widehat{q}_{\alpha\beta}(1-\delta_{\alpha\beta}) \,.
\end{equation}

Within a K-th order Parisi replica breaking scheme we assume 
that the spin overlap matrix has the following (ultrametric) structure
\begin{equation}
\begin{split}
q_{\alpha\beta}=q_i,\quad \text{if}\,\,\,
    I(\alpha/m_i)\neq I(\beta/m_i)\,\,\,\text{and}\,\,\,
              I(\alpha/m_{i+1})= I(\beta/m_{i+1})
\label{parisi-scheme}
\end{split}
\end{equation} 
with $\{q_i\}_{i=0,\ldots,K}$ a set of real numbers and 
$\{ m_i\}_{i=1,\ldots,K}$ a set of integers such that $m_{i+1}/m_i$ is an
integer ($m_0=n$,$m_{K+1}=1$), and we introduce the inverse of the Parisi
function 
\begin{equation}
      x(q)=n+\sum^{K}_{i=0}(m_{i+1}-m_i)\Theta(q-q_i) \,.
\end{equation}
In the limit $K \to \infty$, $q_0\to0$ and $q_K\to q_M$ such that the free
energy per site reads in the limit $n \to 0,\,\,  N \to \infty$
\begin{equation}
\begin{split}
\beta f[x(q),m]=
 &-\frac{\beta}{2}m^2-\frac{u_0\beta }{4}m^4
    +\frac{\alpha\beta}{2}-\Big(\alpha
      +\frac{1}{2}\Big)\ln 2\pi-\frac{1}{2}
       +\frac{\alpha }{2}\Big[\ln\big[1-\beta(1-q_M)\big]
\\& \hspace{-2cm} -\beta\int_0^{q_M}\frac{dq}{1-\beta\int_{q}^1 x(q')dq'}\Big]
    -\frac{1}{2}\Big[\int_{0}^{q_M} \frac{dq}{\int_{q}^1x(q')dq'}
          +\ln(1-q_M)-\frac{m^2}{\int_{0}^1 x(q)dq}\Big] \,,
\end{split}
\end{equation}
where $f$ has to be extremized with respect to $x(q)$ and $m$.
Taking, in general, $x(q)$ to be a piece-wise continuous function it is
straightforward to show following \cite{CS92, N95} that the solution for
$x(q)$ reads
\begin{equation}
    x(q)=\Theta(q-q_{M})
\end{equation}
and, hence, the replica solution is exact. The free energy then reads
\begin{equation}
\begin{split}
\beta f &= \underset{m,q}{\mbox{extr}}
  \left\{-\frac{\beta}{2}m^2-\frac{u_0\beta }{4}m^4+\frac{\alpha\beta}{2}
  \right.
  -\Big(\alpha+\frac{1}{2}\Big)\ln 2\pi-\frac{1}{2}
\\& \qquad  \left.  
 +\frac{\alpha }{2}\Big[\ln[1-\beta(1-q)]-\frac{\beta q}{1-\beta(1-q)}\Big]
 -\frac{1}{2}\Big[\ln[1-q]+\frac{q-m^2}{1-q}
                 \Big]\right\}
\end{split}
\end{equation}
with $q_M=q$, the Edwards-Anderson orderparameter and $m^1=m$.

In agreement with this, one can show by investigating the stability against
replica symmetry breaking fluctuations that the replicon eigenvalue is zero
and, hence, the replica symmetric result is marginally stable. 

The saddle-point equations then become simple algebraic equations for $q$
and $m$ 
\begin{equation}
\begin{split}
   &m\Big(u_0m^2-\frac{1-\chi}{\chi}\Big)=0\,,
            \qquad \frac{q-m^2}{\chi^2}=\frac{\alpha q}{(1-\chi)^2}
\label{statics_equations}
\end{split}
\end{equation}
with $\chi=\beta(1-q)$ the susceptibility. We immediately remark that when
we remove the quartic interaction by setting $u_0=0$ there exist no 
solution with $m \neq 0$ and, hence, there is no retrieval possible.

Henceforth, we take $u_0=1$. The corresponding $T-\alpha$ phase diagram is
shown in Fig.~1.
\begin{figure}[h]
\centering\includegraphics[height=6cm]{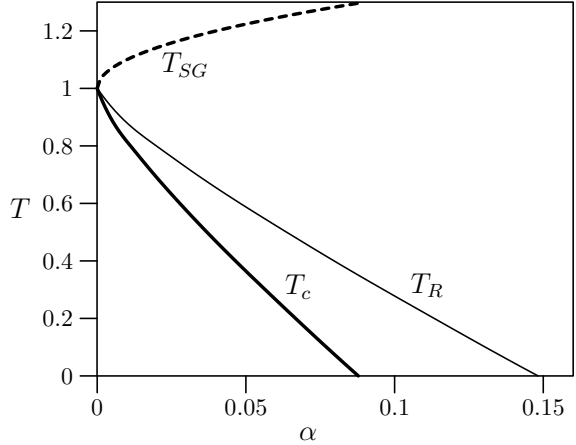}
\caption{\footnotesize{$T-\alpha$ phase diagram for the spherical Hopfield 
model. Full (dashed) lines indicate discontinuous (continuous)
transitions: $T_{SG}$ describes the spin glass transition and $T_R$
(\ref{TR1})-(\ref{TR2})
indicates the border of existense of the retrieval solutions, $T_c$ denotes
the thermodynamic transition below
which the retrieval states are global minima of the free energy.}}
\end{figure}
Let us compare this result with the phase diagram of the standard Hopfield
model calculated in a replica symmetric approximation \cite{HKP91,AGS87}. 
Again we have three phases. For temperatures above the broken line 
$T_{SG}$, there exist paramagnetic solutions characterized by $m=q=0$, 
while below the broken line,  spin glass solutions, $m=0$ but $q\neq0$, 
exist. The transition between the  paramagnetic and the spin glass phase is
continuous and the line $T_{SG}$ separating the two phases is easily 
computed 
\begin{equation}
  T_{SG}(\alpha)=1+\sqrt{\alpha}
\end{equation}
as in the standard Hopfield model. 
Next, below the thin full line $T_R$, (locally stable) retrieval solutions,
$m \neq 0$ and $q\neq0$, appear. The transition between the spin-glass and
the retrieval phase is, however, a discontinous transition.  Using then 
the information that the values of both the order parameters $q$ and $m$
jump when passing this line $T_R$, it is possible to find an analytical 
expression for the latter. Writing the equations \eqref{statics_equations}
as two polynomials in $q$ and $m$, a jump means that complex roots of these
two polynomials become real at the same time for all values of $\alpha,T$
obtained when crossing $T_R$. After some algebra this leads to 
\begin{equation}
\alpha_{R-SG}(\beta)
   =\frac{2}{81\beta^5}\Big[8(\beta-1)^2\beta^2(7\beta-16)
    +\text{sgn}(\Delta)\frac{\Sigma}{\sqrt[3]{|\Delta|}}
          +\text{sgn}(\Delta)\sqrt[3]{|\Delta|}\Big]
\label{TR1}
\end{equation}
with
\begin{equation}
\begin{split}
\Delta&=
 81\sqrt{3}(\beta-1)^{9/2}\beta^{17/2}(|125\beta-128|)^{3/2}-\Delta_2
   \\
\Delta_2&=
  (\beta-1)^5\beta^6
      \{-2097152+\beta(4849664+\beta[-3459072+625\beta(1088+25\beta)])\}
      \\
\Sigma&=
    (\beta-1)^3\beta^4\{-16384+5\beta[6144+\beta(-2976+125\beta)]\}\,.
\end{split}
\label{TR2}
\end{equation}
For $T=0$ this line $T_R$ leads to the critical capacity 
$\alpha(T=0)={4}/{27}=0.14814815$, as seen in Figure 1. 

Below the thick full line, $T_c$, the retrieval states become globally 
stable. 
At this point we remark that, compared with the standard Hopfield model, 
there is no re-entrance behaviour in the transitions since the spherical
model is (marginally) replica symmetric. Furthermore, the region of global
stability for the retrieval solutions in the spherical model is much 
larger.

\section{Macroscopic Dynamics.}

In this section we study the dynamics of the model. Since the neurons are
continuous, we assume the relaxational dynamics to be given by the set of
Langevin equations
\begin{equation}
\frac{\partial \sigma_i (t)}{\partial t}=
   -\mu(t)\sigma_i (t)
     -\frac{\delta {\mathcal H}[\bsigma]}{\delta \sigma_i (t)}
            +\eta_i(t)+ \theta_i(t)
\end{equation}
where the first term on the rhs controls the fluctuations of the neurons,
$\eta_{i}(t)$ is gaussian noise with as first two moments
\begin{equation}
\oter\eta_i(t)\cter=0\,,
     \qquad\oter\eta_i(t)\eta_j(t')\cter=2T\delta_{ij}\delta(t-t')
\end{equation}
and $\theta_i(t)$ is an external perturbation field.
We discuss this dynamics using the generating functional approach 
\cite{MSR73,Do78}. In a straightforward way the following generating 
functional is introduced 
\begin{equation}
{\cal Z}[\bpsi]
  =\int {\mathbf D}[\bsigma,\widehat{\bsigma}]
    \exp\Big[\sum_{i=1}^N\int dt\psi_i(t)i\sigma_i(t)
                             +A[\bsigma,\widehat{\bsigma}]\Big]
\end{equation}
with the action $A[\bsigma,\widehat{\bsigma}]$ given by
\begin{equation}
A[\bsigma,\widehat{\bsigma}]
=\sum_{i=1}^N\int dt\Big\{-T\widehat{\sigma}^2_i(t)
   + i\widehat{\sigma}_i(t)
      \Big(\frac{\partial \sigma_i (t)}{\partial t}
       +\mu(t)\sigma_i (t)-\theta_i(t)
       +\frac{\delta {\cal H}[\bsigma]}{\delta \sigma_i (t)}\Big)\Big\}
\end{equation}
where the  $\psi_i(t)$ are the generating fields, $\widehat{\sigma}_i(t)$
are conjugate variables and ${\mathbf D}[\bsigma,\widehat{\bsigma}]$ is the
measure in path space. From this expression all physical quantities of 
interest can be computed by taking derivatives with respect to the 
generating and the external perturbation fields. 
Assuming that the initial neuron configuration is correlated with only
one pattern, i.e. the condensed one, and averaging over the disorder we
arrive, after some algebra, at the following effective single site 
dynamics in the thermodynamic limit
\begin{equation}
\frac{\partial \sigma(t)}{\partial t}
   =-\mu(t)\sigma(t)+\big[m(t)+u_0m^3(t)\big]\xi^1+\theta(t)
   +\alpha\int dt'[(\unity-{\bf G})^{-1}{\bf G}](t,t')\sigma(t')+\phi(t)
\end{equation}
where $\phi$ is  a colored noise with mean value and variance 
\begin{equation}
\oter\phi(t)\cter_\star=0 \,, \quad 
 \oter\phi(t)\phi(t')\cter_\star=
                   2T\delta(t-t')
    +\alpha[\unity-{\bf G}]^{-1}{\bf C}[\unity-{\bf G}^\dag]^{-1}(t,t')\,.
\end{equation}
In this expression $m(t)$ is the retrieval overlap
\begin{equation}
  m(t)=\odis \xi^1\oter\sigma(t)\ceff\cdis_{\xi^1}
\end{equation}
and the matrices ${\bf C}$ and ${\bf G}$ are the dynamical order parameters
of the problem, i.e. the correlation and response functions
\begin{equation}
C(t,t')=\odis\oter{\sigma}(t){\sigma}(t')\ceff\cdis_{\xi^1}\,,
 \quad 
G(t,t')=\frac{\partial}{\partial \theta(t')}
            \odis\oter\sigma(t)\ceff\cdis_{\xi^1}
\end{equation}
with $\odis\cdots\cdis_{\xi^1}$  the average over the condensed pattern and
$\oter\cdots\ceff$  the average over the effective noise $\phi$.
At this point we remark that in the course of the calculation  we see 
explicitly that the quartic term in the Hamiltonian gives no extensive 
noise contribution. 

Taking the perturbation field $\theta(t)$ to zero, using the causality 
properties of the response function and the spherical constraint, 
$C(t,t)=1$, it is possible to write down a closed set of equations for the
macroscopic observables specifying the dynamics
\begin{equation}
\begin{split}
\Big(\frac{\partial }{\partial t}+\mu(t)\Big)m(t)
  &=\big[m(t)+u_0m^3(t)\big]+\alpha\int_{-\infty}^t dt'R(t,t')m(t')
   \\
\Big(\frac{\partial }{\partial t}+\mu(t)\Big)G(t,t')
  &=\delta(t-t')+\alpha\int_{t'}^{t} dt_1R(t,t_1)G(t_1,t')
    \\
\Big(\frac{\partial }{\partial t}+\mu(t)\Big)C(t,t')
  &=2TG(t',t)+\alpha\int_{-\infty}^{t'} dt_1S(t,t_1)G(t',t_1) 
\\
     &+\big[m(t)+u_0m^3(t)\big]m(t')
     +\alpha\int_{-\infty}^t dt_1R(t,t_1)C(t',t_1)
\label{dynamics_equations}
\end{split}
\end{equation}
where we have defined the  effective correlation function, $S(t,t')$, and
response function, $R(t,t')$, as 
\begin{equation}
\begin{split}
S(t,t')=[\unity-{\bf G}]^{-1}{\bf C}[\unity-{\bf G}^\dag]^{-1}(t,t')\,, 
\quad
R(t,t')=[(\unity-{\bf G})^{-1}{\bf G}](t,t')\, .
\end{split}
\end{equation}

In order to obtain the stationary state from the equations 
(\ref{dynamics_equations}) we assume that, close to equilibrium the
order parameters become time translation invariant meaning that the 
one-time quantities become time independent and the two-time quantities 
satisfy
$C(t,t')= C(t-t'), \,\, G(t,t')= G(t-t')$ (and similarly for $R$ and $S$). 
Time translation invariance holds when the system is ergodic. Then 
the correlation and response functions are related through the 
fluctuation-dissipation theorem (FDT) \cite{Cr93},\cite{CrFe02} 
\begin{equation}
 \beta \partial_{\tau} C(\tau)=G(-\tau)-G(\tau)\, ,
    \quad \beta \partial_{\tau} S(\tau)=R(-\tau)-R(\tau)
\end{equation}
with $\tau=t-t'$, the initial time  $-\infty$ and 
${\partial }/{\partial \tau}$ denoted by $\partial_{\tau}$.
With these assumptions we can write the evolution equation for the 
correlation function in the following way
\begin{eqnarray}
\big(\partial_\tau+\mu-\alpha\beta[1-S(\tau)]\big)C(\tau)
 +\alpha\beta\int_{0}^{\tau} dt'\big[S(\tau-t')-S(\tau)\big]
          \partial_{t'}C(t')\nonumber \\
=\big[m+u_0m^3\big]m 
 +\alpha \int^{\infty}_{0} dt'\big[R(t'+\tau)C(t')+S(t'+\tau)G(t')\big] \,.
\end{eqnarray}
With the conditions $C(0)=1$ and 
$\beta\partial_\tau C(\tau)\big|_{\tau=0}=-1$ we then have for $\tau \to
\infty$
\begin{equation}
\begin{split}
\big[m+u_0m^3\big]m 
 =\frac{C(\infty)}{\beta[1-C(\infty)]}
                    -\alpha\beta S(\infty)[1-C(\infty)] \,.
\end{split}
\end{equation}
Finally, using the evolution equation for $m$ and 
\begin{equation}
\lim_{\tau\to\infty}C(\tau)=q\,, \quad
    \lim_{\tau\to\infty}S(\tau)=\frac{q}{[1-\beta(1-q)]^2}
\end{equation}
it is straightforward to arrive at the equilibrium saddle-point equations
\eqref{statics_equations}. 

At this point we remark that following \cite{Cr93} by starting from the 
evolution equation for the correlation function written as
\begin{equation}
\left(\partial_{\tau}+
\mu-\alpha\beta[1-S(\tau)]\right)\beta[1-C(\tau)]
   -\alpha\beta^2\int_{0}^{\tau} dt'\big[S(\tau-t')
                             -S(\tau)\big]\partial_{t'}C(t')=1
\end{equation}
and using the fact that the dynamics is purely relaxational, and hence, 
that $\partial_t C(t)\leq 0, \,\, \partial_t S(t) \leq 0 $, we can derive 
the following condition for ergodicity
\begin{equation}
\begin{split}
\frac{1}{\beta[1-C(\tau)]}+\alpha\beta-\mu-\mu S(\tau)\geq0 \,.
\end{split}
\end{equation}
This inequality seems to indicate that our system is ergodic in the sense of 
\cite{Cr93}.

\begin{figure}[ht]
\includegraphics[width=.46\textwidth,height=5.5cm]{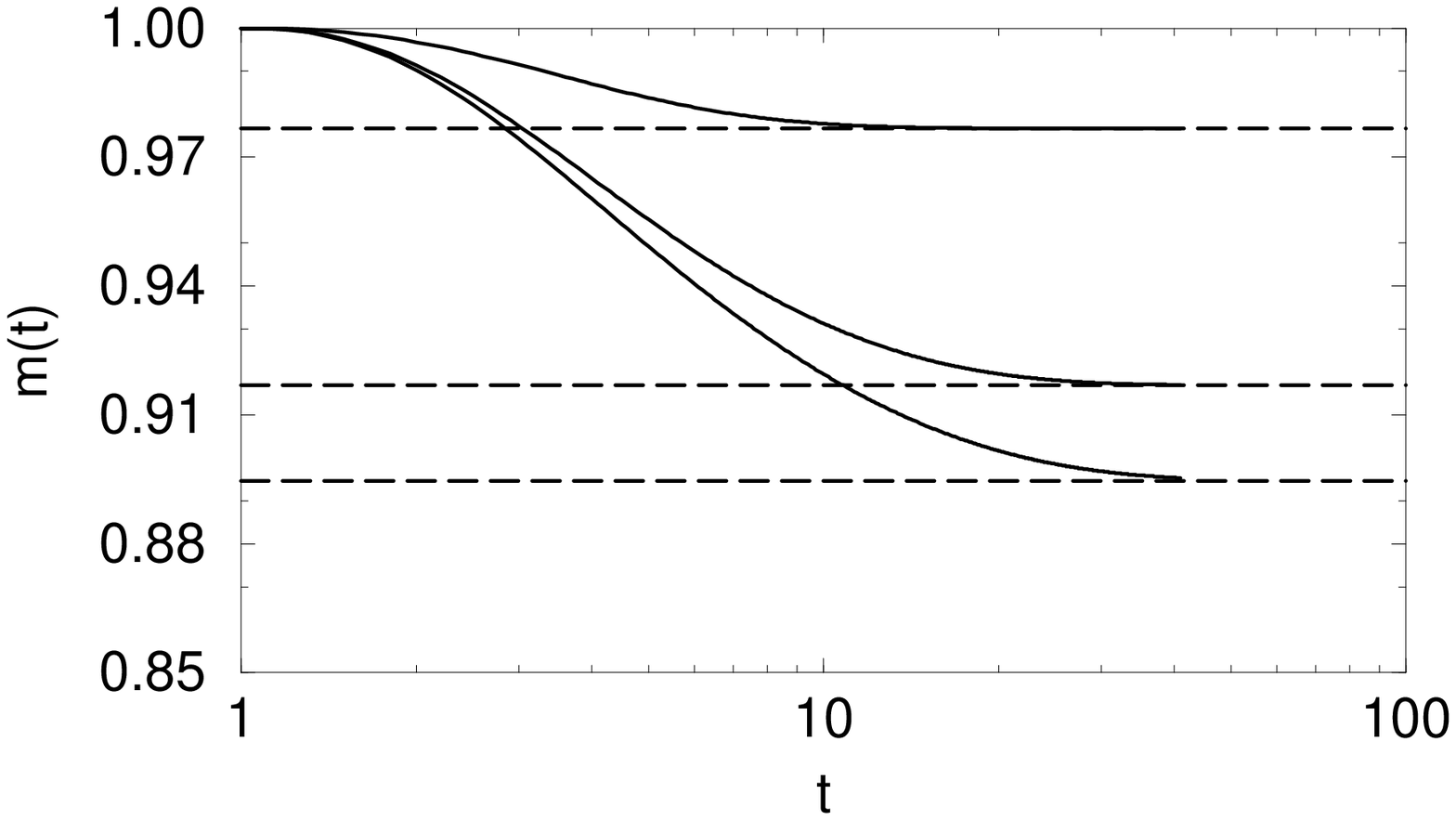}
\hfill
\includegraphics[width=.46\textwidth,height=5.5cm]{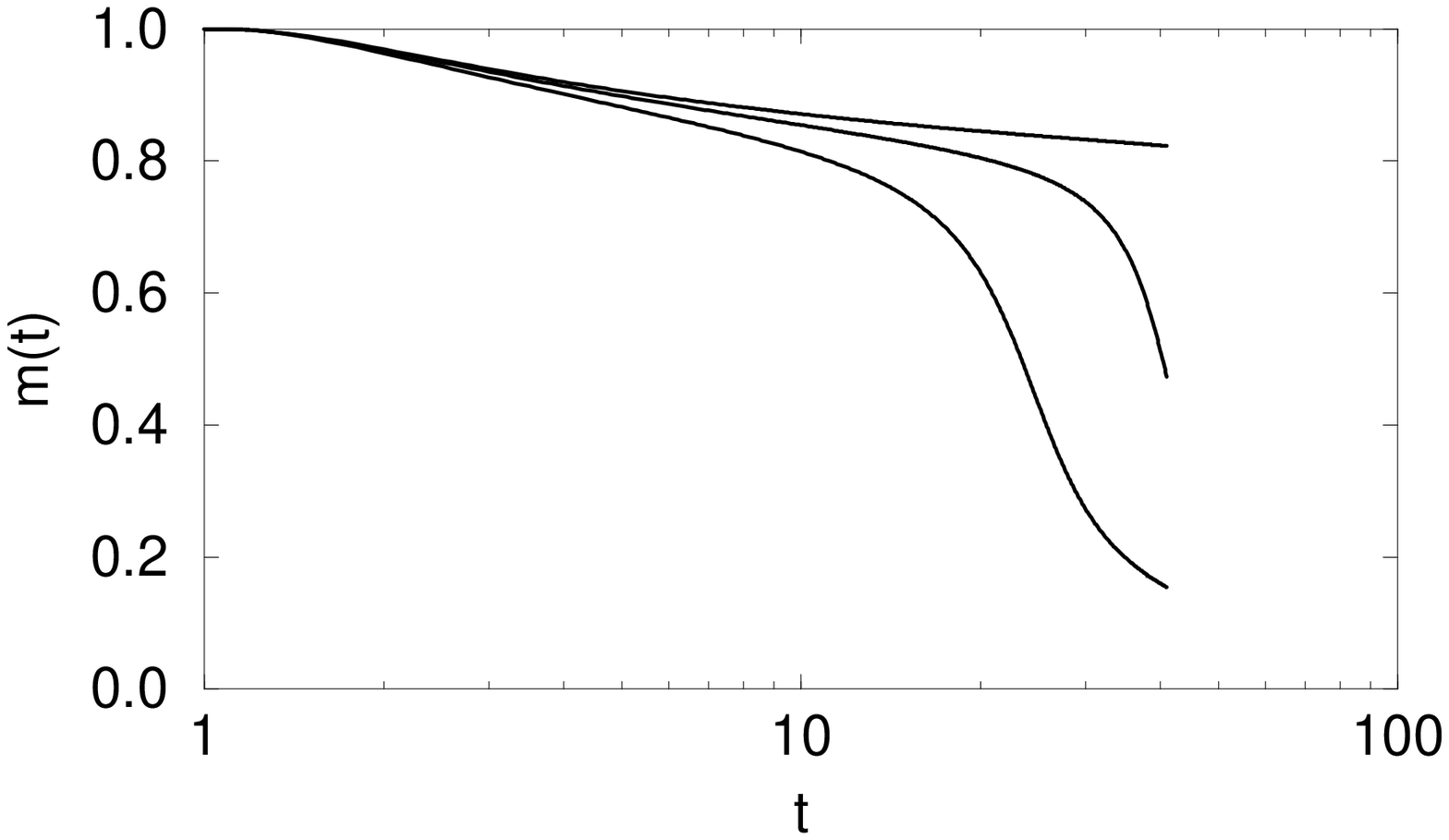}
\label{figure2}
\caption{\footnotesize{The overlap order parameter $m(t)$ as a function of
time at temperature $T=0$ for several values of the capacity: $\alpha= 
0.041,0112,0127$ from top
to bottom on the left and $\alpha=0150,0160,0180$ from top to bottom on the
right. The horizontal dashed lines correspond to the stationary 
value taken from theory.}}
\end{figure}
In fig.~2 we show some plots of the dynamics \eqref{dynamics_equations} 
obtained by  discretizing the equations. Specifically, the overlap order
parameter is shown for several values of the loading capacity at temperature
zero with $m(0)=1$. The behaviour for non-zero temperature is qualitatively 
the same.  We clearly see that below the critical capacity (figure on
the left) the system evolves to the stationary solution, while above the 
critical capacity (figure on the right) $m(t)$ drops to zero.

\section{Conclusions}
In this work we have presented the spherical version of the Hopfield model.
In order to have a retrieval phase a quartic interaction has been 
introduced which does not destroy the spherical character of the model. 
The thermodynamic phase diagram is qualitatively the same as the one for 
the standard Hopfield model, except that there is no re-entrance  because
the system is shown to be (marginally) replica symmetric. Furthermore, the
region of global stability for the retrieval solutions is larger. 
A closed set of equations is obtained for the Langevin dynamics and the 
stationary limit is shown to correspond to the thermodynamic results. A
numerical calculation of the evolution of the retrieval overlap order 
parameter illustrates these findings.  

\section*{Acknowledgments}
We thank  A.C.C. Coolen and R. K\"uhn for informative discussions. This
work has been supported in part by the Fund of Scientific Research,
Flanders-Belgium.

\end{document}